\documentstyle[aps,epsfig,twocolumn]{revtex}

\begin{document}

\title{ Unusual electronic ground state of a prototype
cuprate:\\ band splitting of single CuO$_2$-plane
 Bi$_{2}$Sr$_{2-x}$La$_{x}$CuO$_{6+\delta}$}

\author{C.~Janowitz$^1$, R. M\"uller$^1$, L. Dudy$^1$, A. Krapf$^1$,
 R.~Manzke$^1$, C. Ast$^2$, H. H\"ochst$^2$}

\address{$^1$Humboldt-Universit\"at Berlin, Institut f\"ur Physik, 
Invalidenstr. 110, 10115 Berlin, Germany\\
$^2$Synchrotron Radiation Center, 3731 Schneider Dr.,
53589 Stoughton WI, U.S.A.}

\maketitle

\bibliographystyle{prsty}

\begin{abstract}
The momentum dependence of the split band of Bi$_{2}$Sr$_{1.6}$La$_{0.4}
$CuO$_{6+\delta}$\cite{Ma2001} is investigated by very highly resolved 
photoemission along the $\Gamma$M-line. Since bi-layer effects can not be
present in this single-layer material the results have to be discussed in 
the context of one-particle removal spectral functions. The most prominent 
are microscopic phase separation including striped phase formation, 
critical fluctuations coupled to electrons (hot spots) or even spin charge 
separation within the Luttinger liquid picture, all leading to non-Fermi 
liquid like behavior in the normal state and having severe consequences 
on the way the superconducting state forms.
\end{abstract}

\draft

\begin{figure}[tbp]
  \begin{center}
   \epsfig{file=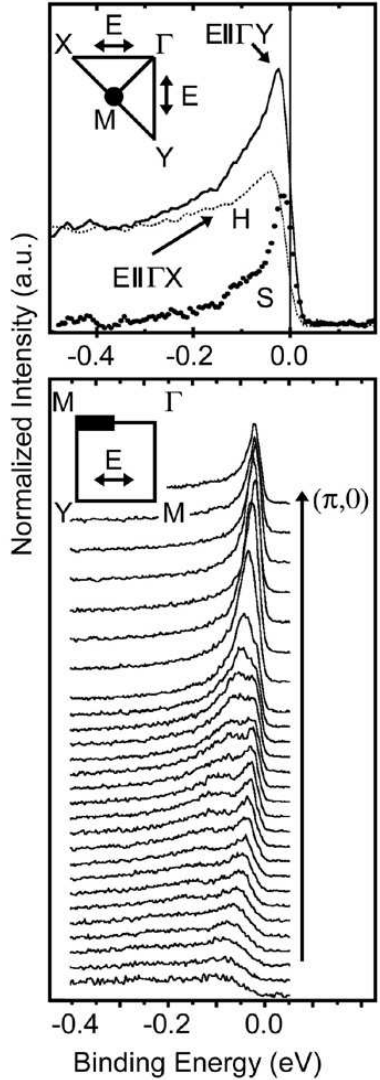,width=0.7\linewidth}
\caption[M-point in-situ polarization dependence for Bi2201]{%
Upper panel:    Normal state (T = 35 K) photoemission spectra of
optimally doped Bi$_2$Sr$_{2-x}$La$_{x}$Cu$_2$O$_{6+\delta}$ 
at the M-point for two polarization geometries as shown in the inset
(h$\nu$=34eV). In this experiment the sample position was 
fixed and the vector potential of the synchrotron radiation
was switched by 90$^{o}$. Drawn line: E $\bot \Gamma$X;
broken line: E$\| \Gamma$X ;
points: difference spectra.\\
Lower panel: Spectra series (T=45K) along a part of the $\Gamma$M- direction 
as indicated in the inset as thick black line with the polarization in
the detection plane (h$\nu$=22eV).}
\label{fig:pol_m}
  \end{center}
\end{figure}

The determination of the {\bf k}-resolved spectral funktion is a key issue
for a deeper understanding of the cuprate superconductors, since even the 
non-superconducting ground state poses severe theoretical challenges
for a discription within some strongly correlated, non-Fermi liquid model. 
The advance achieved over the last years in higher resolved photoemission 
as input for these theories is impressive, but unfortunately to a very great 
extent focussed on Bi$_{2}$Sr$_{2}$CaCu$_2$O$_{8}$, the Bi-cuprate with two 
CuO$_2$-planes per unit cell, where bi-layer splittings are expected to 
influence the photoemission lineshape \cite{Feng01,Chuang01}. 
Therefore, measurements on prototype single CuO$_2$-layer materials like 
Bi$_{2}$Sr$_{2-x}$La$_{x}$CuO$_{6+\delta}$ with sufficiently decoupled layers 
are in high demand to investigate the characteristics of a single CuO$_2$ 
layer slightly above T$_C$ of 29 K with minimum temperature broadening
of the spectra. By resistivity measurements on this compound down
to zero Kelvin in high magnetic fields\cite{Ando96} it could be shown that, 
unlike some other cuprates strict confinement of the carriers to the 
CuO$_2$-planes occurs: it possesses metallic non-Fermi-liquid
character in the planes and insulating character perpendicular to them,
showing a ratio of up to $\rho$$_{c}$/$\rho$$_{ab}$ = 10$^5$ at zero 
Kelvin. On the other hand the observation of two distinct emissions
near the Fermi energy on Bi$_{2}$Sr$_{2-x}$La$_{x}$CuO$_{6+\delta}$
by \it in situ \rm polarization dependend photoemission near the 
($\pi$,0)-point by Manzke et al. \cite{Ma2001} has enormous consequences
on the description of the ground state of cuprate superconductors.
In this letter we will therefore further investigate the individual 
dispersions of these emissions for the n=1-layer Bi-cuprate at optimum 
doping and discuss the observed splittings within the framework of existing 
non-Fermi liquid theories.\\
\\
For photoemission with 34 eV photon energy and \it in situ \rm variation 
of polarization at BESSY in Berlin we used an expermimental setup described 
previously\cite{Ma2001,ja99}. The energy resolution was 30 meV and the angle 
resolution $\pm1$$^o$. The data taken with 22 eV photon energy were obtained 
at the U-PGM beamline of the Synchrotron Radiation Center in Madison-Wisconsin 
(SRC) with a SCIENTA-SES200-analyzer with an overall energy resolution of 
16 meV and the angle resolution $\pm0.18$$^o$. We present photoemission data 
from optimally doped Bi$_2$Sr$_{2-x}$La$_x$CuO$_{6+\delta}$ ( x= 0.40, 
T$_C$ = 29 K ) single crystals. All measurements were 
performed slightly above the critical temperature (35 K at BESSY and 45K at 
SRC). The samples were rectangular shaped with the long side along the 
crystallographic a-axis, confirmed by Laue-diffraction, and have a typical 
size of 5x2 mm$^2$. The growth of the single crystals has been described 
previously \cite{rmue00,ceja99}.\\
\\
FIG.\ref{fig:pol_m} gives a direct indication of the splitting of the Zhang 
Rice singlet state band by different methods. In the upper panel the 
separation of two distinct emissions at the ($\pi$,0)-point of the Brillouin 
zone is demonstrated by applying different polarization geometries. Both 
spectra were taken without moving the analyzer or the sample in succession 
within one hour. The polarization change was solely achieved by tuning the 
crossed undulator beamline. Following this procedure allowed the same sample 
spot to be measured as well as eliminating potential uncertainities due to 
sample misalignment. Although both polarization geometries, described in 
the figure caption of FIG. \ref{fig:pol_m} are identical with respect to dipole
matrix element selection rules, large differences in the intensity
close to E$_F$ occur. Since the spectra are absolutely normalized 
to photon flux the almost vanishing polarization dependence of the
emission at higher binding energy (labelled H) is not an artefact.
The polarization acts strongly at the second spectral contribution
at E$_F$ (labelled S), almost switching it on and off. The difference 
spectrum therefore shows solely emission S, its halfwidth having the size 
of the total energy resolution of 30meV. This sharp emission S will be 
investigated further below. While the spectra in the upper panel were 
recorded with a relatively large angular resolution of $\pm$ 1$^o$, requiring
to work at the M-point where the dispersion of the bands is minimal, in the 
lower panel we show spectra taken at the SRC with an angular resolution of 
$\pm$ 0.18$^o$ and an energy resolution of 16 meV enabling a detection of 
the dispersion of the states. Applied is a geometry in which both emissions, 
S and H, are excited. One sees clearly the dispersion of two individual 
emissions aproaching to a structure not resolvable into individual 
contributions near the the M-point. Additional spectra at higher angles 
(not shown) showed an symmetric reappearance of the two individual structures
around the M-Point. Also at higher photon energy (32eV) two individual
contributions were observable. 

\begin{figure}[tbp]
  \begin{center}
   \epsfig{file=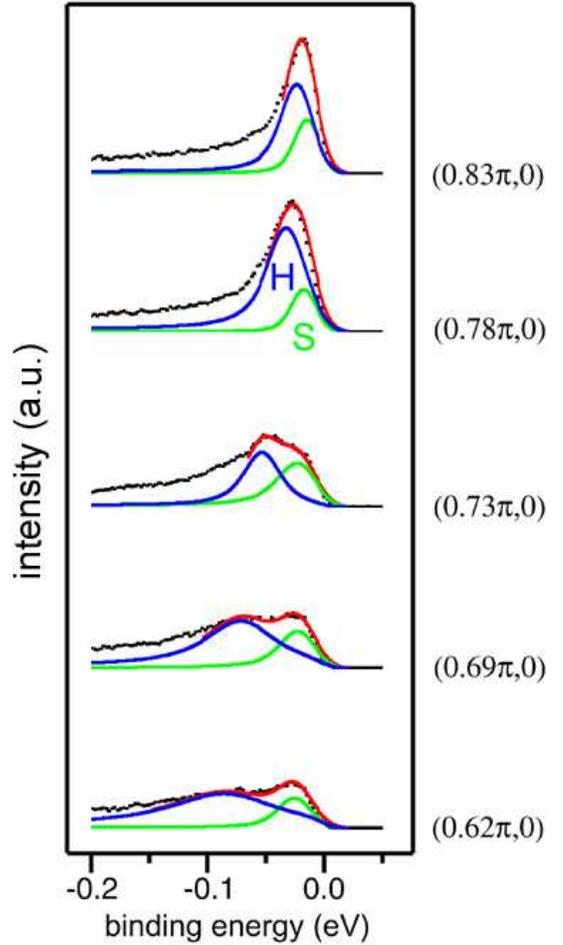,width=0.9\linewidth}
    \caption[EDC spectra along $\Gamma$M for Bi2201]{%
Selection of normal state (T = 45 K) photoemission spectra
of FIG. \ref{fig:pol_m}. The fit is described in the text. 
The shortcoming of the fit towards the higher binding energy 
tail of the spectra is a result of the asymmetry associated 
with the higher binding energy peak which was not considered 
in our simple model calculation.}
   \label{fig:Ga_M}
  \end{center}
\end{figure}

In FIG. \ref{fig:Ga_M} a selection of spectra 
along the $\Gamma$-M direction from FIG. \ref{fig:pol_m} is shown together 
with a fit of the spectral lines contributing to the spectra. The spectra 
were modelled by two Lorentzians convoluted with a Gaussian of appropriate 
width for the experimental resolution and cut off by a Fermi-Dirac 
distribution. The background, although relatively weak in the spectra series, 
was only modelled by the Shirley background\cite{text1}.
Thus, for some spectra it was found necessary to set a limit to the fit region
relatively close to the peak at higher binding energy due to its asymmetric
shape. This emission line H at higher binding energies has a larger halfwidth 
and stronger dispersion relative to the low energy line S. The fit yielded 
typically an intrinsic width of 13 meV for emission S and 51 meV for emission 
H around the M-point. Both are found to be separable by the fit even at the 
M-point. The splitting has been verified on a total of three samples. 

\begin{figure}[tbp]
  \begin{center}
   \epsfig{file=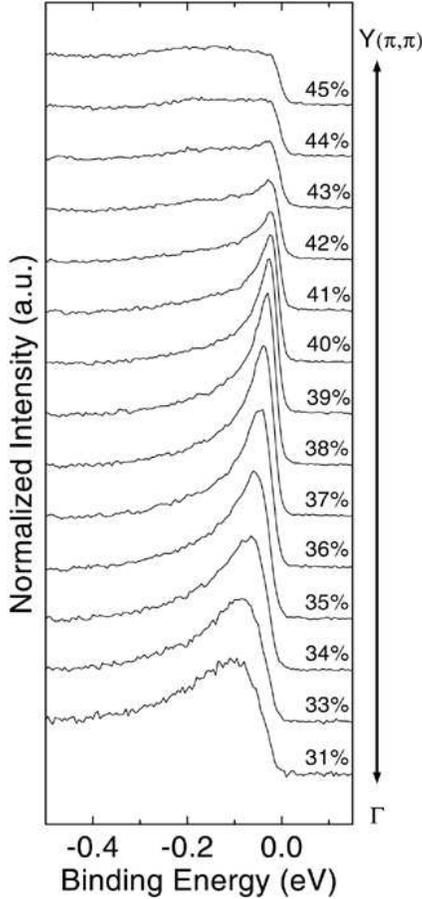,width=0.7\linewidth}
   \caption[EDC-spectra along $\Gamma$Y for Bi2201]{%
Normal state (T = 45 K) photoemission spectra 
in the EDC-mode (energy distribution curve) for optimally doped
Bi$_2$Sr$_{2-x}$La$_{x}$Cu$_2$O$_{6+\delta}$ along $\Gamma$Y. The 
polarization vector lies in the plane of the electron detection, 
the photon energy was 22 eV.}
   \label{fig:Ga_Y}
  \end{center}
\end{figure}

The spectra series along the $\Gamma$-Y direction in FIG. \ref{fig:Ga_Y}
shows the dispersion of the uppermost band in the Fermi surface region 
between the diffraction replicas due to the incommensurate (1x5)-superstructure
(not shown). Unlike for the $\Gamma$M-direction a clear separation into two 
individual contributions is not obvious at any of these spectra. As a test 
these EDC spectra were tentatively fitted similarly as before with two 
Lorentzians. One obtains  two structures that do not merge at the Fermi 
energy crossing point. The dispersion of the low binding energy feature can 
be traced when crossing the Fermi energy. The high binding energy feature 
loses spectral weigth upon approaching the Fermi energy and becomes very broad,
making a definite statement whether it crosses E$_F$ or just approaches 
intimately very hard. One may conclude from this that unlike for the 
$\Gamma$-M direction essentially one asymmetric peak is present along 
$\Gamma$Y.

\begin{figure}[tbp]
  \begin{center}
   \epsfig{file=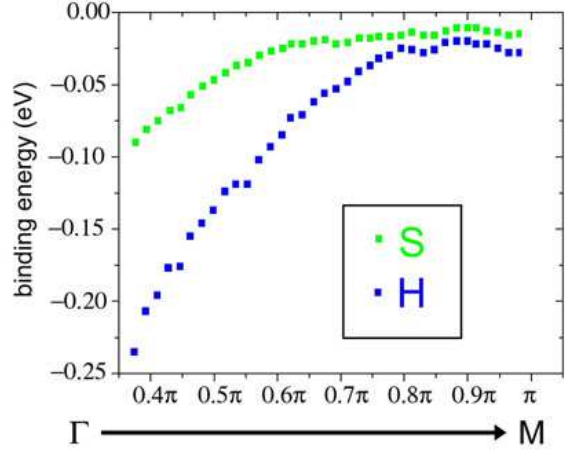,width=1\linewidth}
   \caption[Dispersion of features along $\Gamma$M for Bi2201] {
of states near E$_F$ as derived from FIG. \ref{fig:pol_m}}
   \label{fig:bands}
  \end{center}
\end{figure}

To arrive at a more quantitative understanding of the individual dispersions
the fit results along $\Gamma$-M are shown in the dispersion curve of 
FIG.\ref{fig:bands}. The dispersion of the states can be traced continuously 
with respect to halfwidth and intensity. While the halfwidth of the low 
binding energy peak does not change significantly the higher binding energy 
peak sharpens considerably upon approaching M. If the dispersion of the lower 
binding energy feature is scaled by a factor of approximately 2.7 it coincides 
in dispersion with the higher binding energy feature.

To attribute the observed splitting of the CuO$_2$-derived band
unequivocally to an intrinsic origin, disturbing influences
as resulting from the diffraction replicas or from possibly occuring BiO-pocket
bands must be ruled out. In a detailed study Singh and Pickett\cite{sipi95} 
calculated the near Fermi surface bands and the Fermi surface topolgy by a 
LAPW-method including spin-orbit effects for Bi-2201 with orthorombic
distortion. They found five bands with origin from the BiO-plane  forming
small hole pockets around the M-points and one CuO$_2$-derived band
forming a large Fermi-surface round the X(Y)-points. From our measurements 
along $\Gamma$M depicted in FIG. \ref{fig:pol_m} and FIG. \ref{fig:Ga_M} 
no such E$_F$-crossing of BiO-bands is observed, which should be detectable 
with the high energy and angular resolution applied. This finding is similar 
to the situation for Bi-2212, where also no crossing of BiO-bands was 
reported in this direction. The other source of additional structures, 
namely diffraction replicas can be ruled out from three experimental 
observations: the two bands are degenerate on the entire, main Fermi-surface,
they merge at temperatures exceeding 120 Kelvin and also for increased
hole doping, which brings the system closer to the Fermi-liquid regime.
These issues will be discussed in a forthcoming publication. The possibility 
of surface states can, most probably also be ruled out:
these may be expected in more three-dimensional materials, preferably 
with broken bonds on the surface. Bi-2201 in contrast is extremely 
two-dimensional \cite{Ando96} and cleaves between the weakly bonded 
BiO-layers. No surface reconstruction has been observed so far.
Furthermore, the well studied Bi-2212 is also known to be free of surface
states.\\
\\
In the case of Bi-2212 the dispersion of a single spectral feature near 
E$_F$ has been discussed with the marginal Fermi liquid model of 
Varma \cite{Varma89} or alternative approaches \cite{MaMu99,SheSa99}.
Another basic issue is the dimensionality of the system of interacting 
fermions in the CuO$_2$-planes. Anderson \cite{And90} has put forward 
the tomographic Luttinger model leading to Luttinger liquid behavior with
spin and charge separation. In this context, Castellani et al. 
\cite{Castcast94} have investigated the low energy behavior
of interacting fermions in continuous dimensions between one
and two, discussing the possibility to obtain Luttinger liquid
like behavior in this intermediate dimension. The idea of spin and 
charge separation is also underlying the RVB-theory, first proposed by 
Anderson \cite{Anderson87}. From photoemission-studies \cite{gofron94} 
it has been argued that the extended saddle point in $\Gamma$M-direction 
would render the band structure quasi-1D. Although the cuprates are 
structurally quasi two-dimensional there is increasing theoretical and 
experimental evidence that over a certain range of temperatures on a local 
scale stripe correlations make the electronic structure quasi-one 
dimensional \cite{EFKL00}. Static or dynamic charge inhomogenity has been 
found in several strongly correlated electronic systems and
is not unique to the high temperature superconductors \cite{EFKL00}.
In the cuprates the formation of stripe correlations
can be understood with the concept of frustrated
phase separation \cite{EmKi93}.\\
\\
The basic physics of spin charge separated, essentially one-dimensional
electron liquids has been treated by Voit \cite{Voit96}. In the Luttinger 
liquid picture the dispersion of the spectral features consisting of two
singularities is described by only three parameters: the renormalized 
coupling constants k$_{\nu}$ (${\nu} = {\rho}, {\sigma}$ for charge and spin) 
are the equivalent of the Landau parameters in the case of the Landau Fermi 
liquid. For spin rotation invariance \cite{Voit96} k$_{\sigma}$ is set
to one i.e. this branch is essentially unrenormalized. With the velocities 
v$_{\rho}$ and v$_{\sigma}$ of the charge and spin fluctuations the scenario 
with 0 $\leq$  k$_{\rho}$ $<$ 1 and  v$_{\rho}$ $>$ v$_{\sigma}$ describes 
repulsive and k$_{\rho}$ $>$ 1, v$_{\sigma}$ $>$ v$_{\rho}$ attractive 
interactions. Our measurements give support to a model with repulsive 
interactions (k$_{\rho}$ $<$ 1), generally assumed for HTc's, since 
the dispersion of the holon is larger by a factor of 2.7 (v$_{\rho}$ $>$ 
v$_{\sigma}$). The experimental fact that the overall dispersion 
of the bands is smaller than obtained from single particle bandstructure
calculations is possibly due to some complete renormalization 
of the electron system. One may think for instance of strong electron-phonon 
coupling. The experimentally observed band-splitting and different dispersions
along the $\Gamma$M-symmetry line of Bi-2201 can presumably not be mapped
1:1 onto this essentially one dimensional Luttinger model, but may find
an explanation within an extention to a modified Luttinger model
for the physical system under study. The different halfwidths of S and H can 
possibly be ascribed to a coupling of phonons to the charge but not to the 
spin degrees of freedom.\\
\\ 
In the context of striped phase ordering Salkola et al. \cite{Salko96} 
studied the consequences of disordered charge stripes and antiphase 
spindomains for the properties of high temperature superconductors. The 
orientation of the stripes along the CuO-bonds ($\Gamma$M direction)
leads to a situation, where the fourfold rotational symmetry of the
ideal CuO$_2$-plane is broken through a plane at 45$^{o}$ to the CuO-bond. 
Reflection symmetry and the according selection rules are nevertheless obeyed 
for planes parallel and perpendicular to the stripes. In this view the slight 
orthorhombic distortion is not incorporated. If its effects are regarded as
less important the polarization effects described in FIG. \ref{fig:pol_m} 
and \cite{Ma2001} may in parts be ascribable to this reduced symmetry.
At present there is to our knowledge no theoretical study matching the present 
results perfectly. The theoretical investigation of a quasi one-dimensional 
HT$_C$ with stripes and its spectral function at the Fermi vector by Carlson 
et al. \cite{COKiEm00} shows a broad, incoherent peak above T$_C$, not
in line with our measurments. This is possibly ascribable to the parameters 
used, which can of course greatly influence the lineshape. Also Seibold et al.
\cite{SeiBe00} discussed the occurence of incommensurate CDW's due to stripes 
but found a manifold of bands in the vicinity of E$_F$ also not corroborated 
by our measurments. Alternatively Caprara et al. \cite{Caprara} have studied 
the single-particle spectral properties of a model of coexisting 
antiferromagnetic and incommensurate charge-density-waves coupled to 
electrons in the context of the stripe-quantum-critical-point scenario. 
The deviations from Fermi-liquid behaviour were found near the points of 
the Fermi surface connected by characteristic wave vectors of the critical 
fluctuations, the so called hot spots. The interactions lead to a transfer 
of spectral weight from the quasiparticle peak to an incoherent peak. The 
spectra series calculated from this model along $\Gamma$-M showed however 
a larger separation of the two contributions than found in our experiment, 
possibly due to the parameters used.\\
\\
In summary, we have observed the dispersion of two individual emissions 
with different velocities along the $\Gamma$-M direction for single-layer
Bi-2201. The dispersion and halfwidth could be traced continuously.
In line with polarization dependend measurements the halfwidth of the low 
binding energy emission is 13 meV around the M-point.
Its singular and highly anisotropic character with respect to
polarization is contrasted by the relatively broad (typ. 51meV),
nearly polarization insensitive second emission at higher binding 
energy. Along the $\Gamma$-Y direction no clear splitting is observed.
The observations were discussed in the context of current
theories giving support to the idea of spin and charge separation,
although no perfectly matching theory was found.
When interpreted in the framework of exactly one-dimensional 
models a situation with repulsive interactions is obtained
from a comparison to the experimentally observed dispersions.

\section{Acknowledgement}
We thank Dr. Voit from the University of Bayreuth for stimulating
discussions.
We would like to thank Dr. H. Dwelk, Dr. S. Rogaschewski and D.~Kaiser for 
sample characterization
and preparation. We gratefully acknowledge 
assistance of the staff of BESSY and SRC-Wisconsin.
This work received funding from the
BMBF (BMBF 05 SB8 KH10), the NSF (DMR-0084402) and the
DPG (MA 2371/1).


\begin{thebibliography}{8}

\bibitem{Ma2001}
R. Manzke et al., Phys. Rev. B {\bf 63}, R100504 (2001).

\bibitem{Feng01}
D.L. Feng et al., cond-mat/0102385 (2001).

\bibitem{Chuang01}
Y.-D. Chuang et al.,
cond-mat/0102386 (2001)

\bibitem{Ando96}
Y. Ando et al.,
Phys. Rev. Lett. {\bf77}, 2065 (1996)

\bibitem{ja99}
C. Janowitz et al.,
J. Electron Spec. Rel. Phen.{\bf 105}, 43 (1999).

\bibitem{rmue00}
R. M\"uller et al.,
Physica C {\bf 341-348}, 2109 (2000).

\bibitem{ceja99}
C. Janowitz et al., Physica B {\bf 259-261}, 1134 (1999).

\bibitem{text1}
It is the intention to apply a very simple
fit procedure to the spectra in order not to interprete the spectral line
physically when, on the other hand, a theoretical description is lacking.

\bibitem{sipi95}
D.J. Singh and W.E. Pickett,
Phys. Rev. B {\bf 51}, 3128 (1995).

\bibitem{Varma89}
C.M. Varma et al., Phys. Rev. Lett. {\bf 63}, 1996 (1989).

\bibitem{MaMu99}
K. Matho, A. M\"uller, Physica C {\bf 318}, 585 (1999).

\bibitem{SheSa99}
Z.-X. Shen, G.A. Sawatzky, Phys. Stat. Sol. (B) {\bf 215}, 523 (1999).

\bibitem{And90}
P.W. Anderson, Phys. Rev. Lett. {\bf 64}, 1839 (1990);
{\bf 65}, 2306 (1990).

\bibitem{Castcast94}
C. Castellani, C. Di Castro, W. Metzner, 
Phys. Rev. Lett. {\bf 72}, 316 (1994).

\bibitem{Anderson87}
P. W. Anderson, Science {\bf 235}, 1196 (1987).

\bibitem{gofron94}
K. Gofron et al., Phys. Rev. Lett. {\bf73}, 3302 (1994)

\bibitem{EFKL00}
V.J. Emery et al.
and references therein, Phys. Rev. Lett.{\bf85}, 2160 (2000)

\bibitem{EmKi93}
V.J. Emery, S.A. Kivelson, Physica {\bf209C}, 597 (1993)

\bibitem{Voit96}
J. Voit, Rep. Prog. Phys. {\bf58}, 977 (1995)

\bibitem{Salko96}
M.I. Salkola, V.J. Emery, S.A. Kivelson,
Phys. Rev. Lett. {\bf77}, 155 (1996)

\bibitem{COKiEm00}
E.W. Carlson et al., Phys. Rev. B {\bf62}, 3422 {2000}

\bibitem{SeiBe00}
G. Seibold et al., Eur. Phys. J. B {\bf13}, 87 (2000)

\bibitem{Caprara}
S. Caprara et al., Phys. Rev. B {\bf59}, 14980 (1999)

\end{thebibliography}
\end{document}